%% file: a_our_work.tex
\documentclass[acmsmall,screen]{acmart}

\AtBeginDocument{%
  }

\usepackage{amsmath,amssymb,amsfonts}
\usepackage{algorithmic}
\usepackage{graphicx}
\usepackage{textcomp}
\usepackage{xcolor}
\usepackage{multirow}
\usepackage{booktabs}
\usepackage{tcolorbox}
\usepackage{diagbox}
\usepackage{tabularx}
\usepackage{microtype}
\usepackage{stackengine}  
\usepackage{subcaption} 
\usepackage{wrapfig}

\usepackage{amsmath}
\usepackage{listings}

\newcommand{\findingboxx}[1]{
\begin{center}
\begin{tcolorbox}[colback=gray!11,
                  colframe=black,
                  boxrule=0.2mm,
                  width=0.9\textwidth,
                  arc=.8mm, auto outer arc,
                 ]
 #1
\end{tcolorbox}
\end{center}}

\newcommand{\tool}[1]{\textcolor{red}{{\it BBC}}}
\newcommand{\Todo}[1]{\textcolor{red}{{\it TODO}}}

\begin{document}

\title{DELA: A Novel Approach for Detecting Errors Induced by Large Atomic Condition Numbers}

\author{Youshuai Tan}
\authornote{Both authors contributed equally to this research.}
\email{tanyoushuai123@gmail.com}
\author{Zhanwei Zhang}
\authornotemark[1]
\email{itbill8888@gmail.com}
\affiliation{%
  \institution{Wuhan University}
    \country{China}
}

\author{Jinfu Chen}
\authornote{Corresponding authors.}
\affiliation{%
  \institution{Wuhan University}
    \country{China}
}
\email{jinfuchen@whu.edu.cn}

\author{Zishuo Ding}
\affiliation{%
  \institution{The Hong Kong University of Science and Technology (Guangzhou)}
    \country{China}
}
\email{zishuoding@hkust-gz.edu.cn}

\author{Jifeng Xuan}
\authornotemark[2]
\affiliation{%
  \institution{Wuhan University}
    \country{China}
}
\email{jxuan@whu.edu.cn}

\author{Weiyi Shang}
\affiliation{%
  \institution{University of Waterloo}
    \country{Canada}
}
\email{wshang@uwaterloo.ca}

\renewcommand{\shortauthors}{Tan et al.}

\begin{abstract}
Numerical programs form the foundation of modern science and engineering, providing essential solutions to complex mathematical problems.
Therefore, errors in numerical results would lead to harmful consequences, especially in safety-critical applications. Since only a few inputs may lead to substantial errors for numerical programs, it is essential to determine whether a given input could result in a significant error. Existing researchers tend to use the results of high-precision programs to assess whether there is a substantial error, which introduces three main challenges: difficulty of implementation, existence of potential faults in the detection of numerical errors, and long execution time.

To address these limitations, we propose a novel approach named DELA. Our approach is based on the observation that most numerical errors stem from large condition numbers in atomic operations (such as addition and subtraction), which then propagate and accumulate. DELA injects small perturbations into the results of individual atomic operations within the program and compares the outcomes of the original program with the perturbed version to detect errors. 
We evaluate DELA with datasets from ATOMU and HSED, as well as data from a complex linear system-solving program. Experimental results demonstrate that we can detect all the significant errors that were reported by prior research. DELA shows strong alignment with high-precision programs of ATOMU and HSED, with average Pearson and Spearman correlations of 0.86 and 0.61. Additionally, DELA effectively detects significant errors in complex programs, achieving correlation scores of 0.9763 and 0.8993. More importantly, in experiments with ATOMU and HSED, DELA’s perturbed programs run within only 0.13\% of the time needed by high-precision versions; while for the linear system-solving programs, DELA is 73.46 times faster than the high-precision programs.

\end{abstract}




\maketitle

\input{intro}
\input{background}

\input{motivation}
\input{definition}

\input{methodology}
\input{evaluation}
\input{discussion}
\input{threat}
\input{relatedwork}

\section{Conclusion}
\label{conclusion}

Existing approaches typically employ high-precision programs to obtain oracles for detecting numerical errors. However, the conventional solutions present three key challenges: implementation complexity, existence of potential faults in the detection of numerical errors, and long execution times. To address these issues, we propose a novel approach, termed DELA, which efficiently and effectively identifies numerical errors. Our experimental results indicate that DELA can tackle these three challenges. In future work, we intend to apply our approach to more practical and complex programs, such as numerical simulation applications, to assist scientists in assessing whether specific inputs could lead to significant errors.

\section{Data Availability}
\label{data}
The replication package, including the datasets and scripts, is available on the link\footnote{https://anonymous.4open.science/r/anonymity-1D01}.

\bibliographystyle{ACM-Reference-Format}
\bibliography{sample-base}

\end{document}

%% file: intro.tex
\section{Introduction}
\label{Introduction}

Numerical programs are crucial to the advancement of science and engineering, powering applications in fields such as aerospace, finance, and data analysis~\cite{he2020testing, xu1994numerical, kotrla1996numerical, orszag1974numerical, rives1992joint, kobayashi1993modeling}. However, these programs inherently depend on floating-point arithmetic, which can have numerical errors, i.e., the difference between the exact mathematical value and the computed approximation of that value due to limitations in a computational process. Such errors in critical systems like rocket launches and market fluctuations in financial applications can have profound impacts \cite{zhang2024hierarchical}.

To address the potential impact of numerical errors, previous researchers have developed various tools aimed at detecting inaccuracies in floating-point computations~\cite{benz2012dynamic, chowdhary2021parallel, chowdhary2022fast}. A typical common approach in numerical error detection relies on high-precision computation programs as oracles to verify the results of lower-precision calculations. By comparing outputs from standard-precision computations against those from high-precision versions using identical inputs, tools can identify discrepancies indicative of errors~\cite{chiang2014efficient, zou2015genetic, yi2019efficient, guo2020efficient, wang2022detecting, zhang2023eiffel, zhang2024hierarchical}. 
While this method can be effective, it also presents several practical limitations.

High-precision programs present three main drawbacks that hinder their adoption: \textbf{1) difficulty of implementation:} High-precision program implementation often requires tailored operations that are specifically designed for various levels of precision, rather than a straightforward substitution of low-precision values with high-precision ones \cite{wang2016detecting, zou2019detecting}. \textbf{2) existence of potential faults in the detection of numerical errors:} 
High-precision programs may harbor the same types of errors as their lower-precision counterparts, leading to a risk of overlooking certain numerical inaccuracies.
\textbf{3) long execution time:} The computational overhead associated with high-precision programs can lead to substantially longer execution times, limiting their practicality in time-sensitive applications \cite{larsson2013exploring}.

To overcome these challenges, we propose a novel approach, named DELA (\underline{D}etecting \underline{E}rrors induced by \underline{L}arge \underline{A}tomic condition numbers), to detect numerical errors. DELA is grounded in the theory that most numerical errors stem from large atomic condition numbers. By injecting small perturbations into atomic operations, DELA can effectively expose latent bugs. 
In response to the implementation challenges associated with high-precision oracles, DELA is designed as an LLVM pass, enhancing ease of use. To evaluate DELA, we utilize the data and code from ATOMU \cite{zou2019detecting} and HSED \cite{zhang2024hierarchical} and we further assess DELA with a linear system-solving program as a complex evaluation case. Our experimental results demonstrate that DELA can detect all the significant numerical errors that were reported by prior research, showing a high correlation with the errors identified by high-precision programs. Additionally, DELA overcomes faults of high-precision programs in the detection of numerical errors and operates much faster than high-precision approaches.

In summary, our work makes the following three main contributions:
\begin{itemize}
\item We introduce DELA, a novel approach designed to efficiently detect numerical errors in floating-point computations.
\item We propose a robust perturbation strategy to maximize the likelihood of uncovering numerical errors.
\item We evaluate DELA on a diverse set of benchmarks, including both popular, i.e., ATOMU and HSED, and complex numerical programs.

\end{itemize}

\textbf{Paper organization.} The remainder of the paper is organized as follows. Section~\ref{bac} presents the background of our work. We introduce the three challenges of the use of high-precision programs as oracles in Section~\ref{motivate}. Section~\ref{definition} explains the novelty analysis of DELA and the problem definition. Section~\ref{method} elaborates on DELA. The evaluation of our approach is presented in Section~\ref{eva}. Section~\ref{discussion} discusses the set of perturbations and practical utility. We discuss limitations and introduce related work in Section~\ref{threat} and Section~\ref{related-work}, respectively. Finally, section~\ref{conclusion} concludes our work and Section~\ref{data} provides a replication package.

%% file: background.tex
\section{Background}
\label{bac}
In this section, we provide an overview of the background knowledge related to numerical error measurement, condition number, and matrix condition number. Such knowledge is essential for understanding the challenges posed by numerical errors, particularly those influenced by large atomic condition numbers.

\subsection{Numerical Error Measurement}
Floating point numbers are the standard numerical representation used in modern numerical programs to simulate real numbers. However, due to the inherent approximation and errors of floating point arithmetic \cite{jezequel2008cadna}, results produced by numerical programs are prone to errors compared to the outcome of mathematical operations over the field of real numbers. For instance, in Java SE Development Kit 21.0.3, subtracting the floating point number 1.1 from 1.2 does not yield 0.1; instead, it results in 0.09999999999999987. In this context, we denote the real number as \(x\) and its floating point representation as \(\hat{x}\). With respect to a floating program \(\textbf{P: }\hat{y}=\hat{f}(\mathbf{\hat{x}})\), \(\hat{y}\) refers to the numerical result and \(f(\mathbf{x})\) represents for the real outcome. The error  between the ideal mathematical result \(f(\mathbf{x})\) and the outcome of the floating point program \(\hat{f}(\mathbf{\hat{x}})\) can be quantified using the following two common metrics:

\begin{equation}
Err_{abs}(f(\mathbf{x}),\hat{f}(\mathbf{\hat{x}}))=\left|f(\mathbf{x})-\hat{f}(\mathbf{\hat{x}})\right|\quad Err_{rel}(f(\mathbf{x}),\hat{f}(\mathbf{\hat{x}}))=\left|\frac{f(\mathbf{x})-\hat{f}(\mathbf{\hat{x}})}{f(\mathbf{x})}\right|
\end{equation}
where \(Err_{abs}\) and \( Err_{rel}\) stand for absolute error and relative error, respectively. However, if the value of \(f(x)\) approaches 0, it can lead to a division-by-zero error. 

To mitigate this issue, Unit in the Last Place \((ULP)\) is often used. 
\(ULP(\mathbf{x})\) is the gap between a given floating point number \(x\) and the next representable floating point number \cite{overton2001numerical}. The formulation for \(ULP\) is given by:

\begin{equation}
ULP(\mathbf{x})=\epsilon\times2^E
\end{equation}
where \(\epsilon\) is the machine epsilon and \(E\) represents for the exponent of \(x\). Therefore, different numbers possess varying \(ULP\) and \( Err_{ulp}\) is flexible:

\begin{equation}
Err_{ulp}(f(\mathbf{x}),\hat{f}(\mathbf{\hat{x}}))=\left|\frac{f(\mathbf{x})-\hat{f}(\mathbf{\hat{x}})}{ULP(f(\mathbf{x}))}\right|
\end{equation}
Therefore, in this work, we use \(Err_{ulp}\) as our primary error measurement metric.

\subsection{Condition Number}
Condition number serves as a crucial indicator of how the relative rounding error in the input \(x\) is amplified during 
the evaluation of \(f(\mathbf{x})\) \cite{overton2001numerical}. Its derivation can be elucidated using the Taylor Expansion Theorem \cite{zou2019detecting}. The numerical error in evaluating \(f(\mathbf{x})\) at a perturbed value \(x + \Delta x\) can be expressed as:

\begin{equation}\begin{aligned}
Err_{rel}(f(x),f(x+\Delta x))& =\left|\frac{f(x+\Delta x)-f(x)}{f(x)}\right| \\
&=\left|\frac{f(x+\Delta x)-f(x)}{\Delta x}\cdot\frac{\Delta x}{f(x)}\right| \\
&=\left|(f^{\prime}(x)+\frac{f^{\prime\prime}(x+\theta\Delta x)}{2!}\Delta x)\cdot\frac{\Delta x}{f(x)}\right|,\theta\in(0,1) \\
&=\left|\frac{\Delta x}{x}\right|\cdot\left|\frac{xf^{\prime}(x)}{f(x)}\right|+O\big((\Delta x)^{2}\big) \\
&=Err_{rel}(x,x+\Delta x)\cdot\left|\frac{xf^{\prime}(x)}{f(x)}\right|+O\big((\Delta x)^2\big)
\end{aligned}
\end{equation}
where \(\Delta x\) represents for small perturbation and \(\theta\) is the Lagrange form of the remainder. Consequently, the equation derives the formula for the condition number \cite{higham2002accuracy}, \(\left|\frac{xf^{\prime}(x)}{f(x)}\right|\). Large condition number would lead to ill-conditioned problems.

Since we can ignore the term \(O\big((\Delta x)^2\big)\), we could conclude that a minor error of the input could result in a significant error in the output if the condition number is large. Additionally, condition number contains the derivative of the function \(f\) with respect to \(x\). As a result, calculating the condition number can be impractical for complex programs \cite{fu2015automated}.

\subsection{Matrix Condition Number}
In the context of linear algebra, the matrix condition number is a fundamental concept. It quantifies the sensitivity of the solution of a linear system of equations to small changes in the coefficient matrix or the right-hand side vector. The number \(\left\|A\right\|\left\|A^{-1}\right\|\) is the matrix condition number of the matrix \(A\), denoted by \(\kappa \left(A\right)\). If \(\kappa \left(A\right)\) is large, achieving an accurate solution for a linear system becomes increasing difficult \cite{ford2014numerical}. 

For instance, considering the matrix \(A=\begin{bmatrix}1.0001&1\\1&1\end{bmatrix}\), this matrix has a large condition number. The large condition number indicates that the linear system \(Ax=b\) is ill-conditioned. A small perturbation in the right-hand side vector \(b\) can lead to a significant change in the solution vector \(x\).
Specifically, the solution of the linear systems \(Ax\)=\(\left[{\begin{array}{cc}1.0001&1\\1&1\end{array}}\right]\left[{\begin{array}{c}x_{1}\\x_{2}\end{array}}\right]=\left[{\begin{array}{cc}2.0001\\2\end{array}}\right]\) and \(Ax\)=\(\left[\begin{array}{cc}1.0001&1\\1&1\end{array}\right]\left[\begin{array}{c}x_{1}\\x_{2}\end{array}\right]=\left[\begin{array}{c}2\\2\end{array}\right]\) yield outputs that differ significantly, despite the small perturbation in \(b\). 
In the fist case, we find \(x_1 = 1\) and \(x_2 = 1\), while in the latter, \(x_1=0.0000\) and \(x_2=2.0000\). This example illustrates how even minimal changes in input can lead to substantial variations in the output when dealing with matrices that possess a high condition number. We take linear system-solving programs as complex programs to evaluate DELA.


%% file: motivation.tex
\section{A Preliminary Study}
\label{motivate}
In this section, we present a study that highlights three significant drawbacks associated with the use of high-precision oracles in numerical error detection. These limitations can adversely affect the reliability of numerical programs.

\noindent\textbf{Motivation:} Many techniques \cite{chiang2014efficient, zou2015genetic, zou2019detecting, yi2019efficient, guo2020efficient, wang2022detecting, zhang2023eiffel, zhang2024hierarchical} have been proposed to identify inputs that could lead to significant errors in numerical programs. These techniques rely on search-based algorithms, which necessitate oracles for guidance. Most of the existing tools utilize high-precision programs to obtain oracles. However, the use of high-precision operations contain three inherent drawbacks: 

\begin{itemize}
    \item \textbf{Difficulty of implementation.} Developing high-precision programs requires specialized knowledge of mathematics and numerical analysis. High-precision computations often involve precision-specific operations rather than merely substituting low-precision numbers with high-precision ones \cite{wang2016detecting, zou2019detecting}. 
    \item \textbf{Existence of potential faults.} When we use high-precision programs to obtain an oracle and calculate numerical errors, it may lead to mistakes, such as overlooking significant errors or incorrectly identifying errors.
    \item \textbf{Long execution time.} In terms of runtime performance, quadruple precision (128 bits) is about 100 times slower than double precision (64 bits) \cite{larsson2013exploring}. Additionally, programs that use arbitrary precision libraries, such as MPFR \cite{fousse2007mpfr} and \textit{mpmath} \cite{mpmath}, face even greater slowdowns as they increase precision levels. 
\end{itemize}

Given these challenges, we conduct a preliminary study to identify and examine these problems, aiming to develop solutions that mitigate their impact on numerical program testing.

\noindent\textbf{Approach:} 
We follow four steps to thoroughly investigate these three challenges.

\noindent\textbf{Step1. Data Selection.} We use the dataset from prior study \cite{zou2019detecting}, which includes 88 functions from the GNU Scientific Library (GSL)\footnote{https://www.gnu.org/software/gsl/}. The GNU Scientific Library (GSL) is a numerical library for C and C++ programmers. It is free software under the GNU General Public License. This GSL dataset provides a diverse range of numerical functions with varying levels of complexity.

\noindent\textbf{Step2. High-precision program implementation.} 
We reuse the implementation from prior study \cite{zou2019detecting}. The authors provide high-precision versions of these GSL functions, implemented using \textit{mpmath}. Given the authors' expertise in this field, these high-precision programs are considered credible and reliable. To illustrate the complexities associated with high-precision program implementation, we apply \textit{cloc}, a tool that quantifies the number of lines of source code for the GSL functions and their \textit{mpmath} counterparts.
We also analyze key branch keywords to identify branching statements, including ``if'', ``else'', ``switch'', and ``case''. Furthermore, we adopt a recursive methodology to traverse function calls up to a specified maximum depth, enabling us to aggregate branch counts across nested function calls.

\noindent\textbf{Step3. Error analysis.}  We analyze the outputs of the original GSL functions and their high-precision counterparts for a wide range of input floating point values. Due to the vast number of floating point numbers, examining the results for all possible inputs is impractical. Therefore, we focus on the recorded inputs from ATOMU that result in significant errors. Additionally, we examine the surrounding points to assess the accuracy of the high-precision implementations.

\noindent\textbf{Step4. Performance evaluation.} We conduct a comprehensive analysis of the runtime and accuracy of both the original GSL functions and their high-precision counterparts. We execute both primitive and high-precision programs of all the 88 functions. Due to the numerous branches within GSL functions, we select 1,000 random input numbers and run all 176 programs, aiming to trigger most possible branches. The input values are restricted to the general interval of [-100, 100], as suggested by prior study~\cite{zhang2024hierarchical}.

\noindent\textbf{Result. Implementing high-precision programs is a demanding and complex task.} 
Developing high-precision programs is a challenging task that goes far beyond simply substituting double-precision floating-point numbers with higher-precision equivalents. Through an analysis of 88 high-precision programs, we observe that these conversions often involve significant changes to branching logic, approximation algorithms, and even parameter tuning. 
Specifically, the average number of lines of code for GSL functions is 162, while for \textit{mpmath} implementations, it is 3,742. Additionally, the average number of branches in GSL is 29 compared to 1,012 in \textit{mpmath}.
This complexity highlights the challenge of manual high-precision code adaptation, which can introduce potential errors if not carefully handled. In our investigation, we discover a notable error within a high-precision implementation of the \textit{gsl\_sf\_hazard\_e} function.

\begin{lstlisting}[language=python, caption=gsl\_sf\_hazard\_e]
                mpmath.npdf(x)/(1-mpmath.ncdf(x))
\end{lstlisting}

where \textit{mpmath.ncdf()} is used to calculate the Cumulative Distribution Function (CDF) of the standard normal distribution. As the input moves towards positive infinity, the denominator approaches zero. We find the result becomes \textit{inf} if the input is bigger than 13.169554. In contrast, the original GSL implementation avoids this issue due to its mathematical approximations. Therefore, we believe this is a bug in the authors' high-precision program implementation. Even experts in the field can make mistakes, demonstrating that converting standard programs into high-precision ones is an extremely complex task.

\begin{wraptable}{r}{0.5\textwidth}
\centering
\caption{\textit{gsl\_sf\_legendre\_Q0\_e} results when the input approaches 1.}
\label{sample_82}
\resizebox{0.48\textwidth}{!}{
\begin{tabular}{lll}
\toprule
Input              & GSL Result         & \textit{mpmath} Result      \\
\midrule
0.9999999999999992 & 17.742018800590866 & 17.742018800590866 \\
0.9999999999999993 & 17.819094140504497 & 17.819094140504497 \\
0.9999999999999994 & 17.910254918901472 & 17.910254918901472 \\
0.9999999999999996 & 18.021826694558577 & 18.021826694558577 \\
0.9999999999999997 & 18.165667730784467 & 18.165667730784467 \\
0.9999999999999998 & 18.36840028483855  & 18.36840028483855  \\
0.9999999999999999 & 18.714973875118524 & 18.714973875118524 \\
1.0                & inf                & nan                \\
1.0000000000000002 & 18.36840028483855  & 18.36840028483855  \\
1.0000000000000004 & 18.021826694558577 & 18.021826694558577 \\
1.0000000000000007 & 17.819094140504497 & 17.819094140504497 \\
1.0000000000000009 & 17.675253104278607 & 17.675253104278607 \\
1.000000000000001  & 17.563681328621502 & 17.563681328621502 \\
1.0000000000000013 & 17.472520550224523 & 17.472520550224523\\
\bottomrule
\end{tabular}
}
\end{wraptable}

\begin{figure}[htbp]
    \centering
    \begin{minipage}{0.45\textwidth} 
        \centering
        \includegraphics[width=\textwidth]{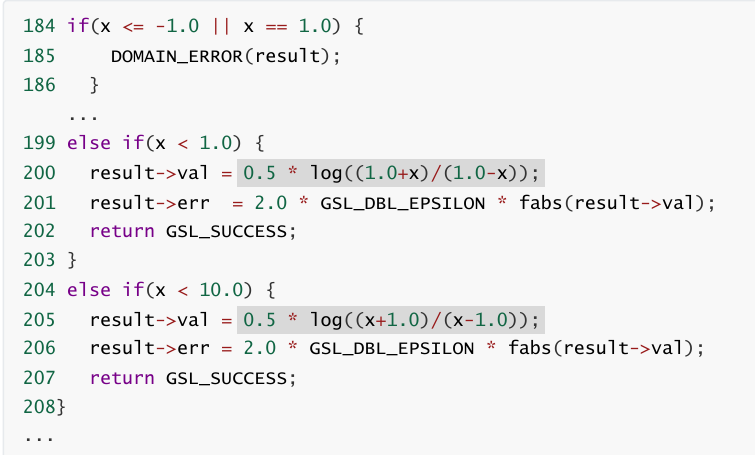}
        \subcaption{Code snippet.}\label{subfigure1}
    \end{minipage}
    \hfill 
    \begin{minipage}{0.5\textwidth} 
        \centering
        \includegraphics[width=\textwidth]{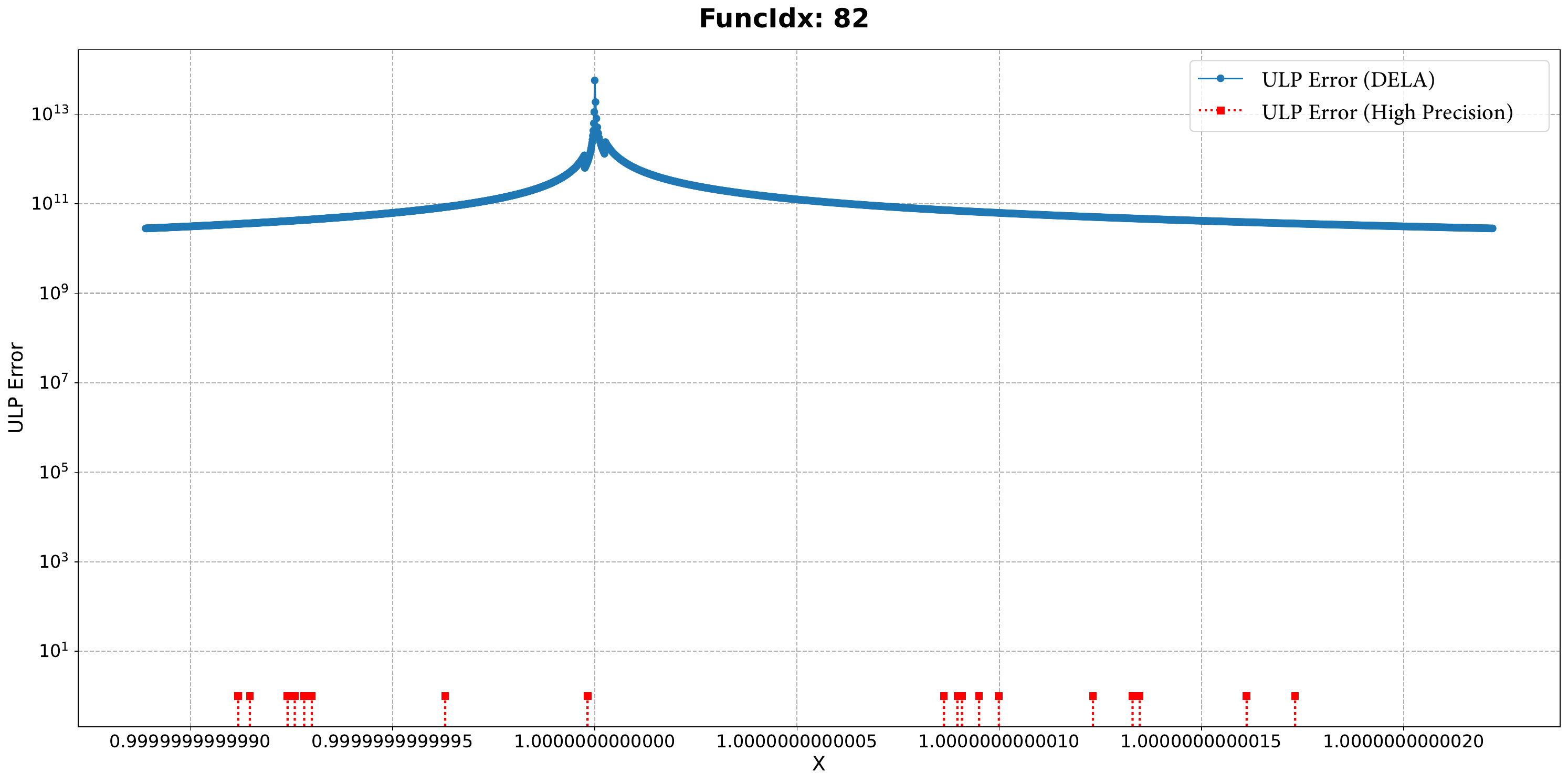}
        \subcaption{Errors near 1 calculated by our approach and high-precision programs.}\label{subfigure2}
    \end{minipage}
    \caption{An illustrative example of the GSL function \textit{gsl\_sf\_legendre\_Q0\_e.}} 
    \label{func82}
\end{figure}

\textbf{Faults in high-precision programs can lead to significant errors being overlooked.} We observe that the high-precision implementation of several GSL functions failed to identify a significant error caused by catastrophic cancellation. Catastrophic cancellation occurs in numerical computations when subtracting two nearly equal numbers results in a significant loss of precision. For example, Figure~\ref{subfigure1} illustrates a code snippet of one GSL function, named \textit{gsl\_sf\_legendre\_Q0\_e}, which highlights a catastrophic cancellation. As \(x\) approaches 1, 1.0-\(x\) becomes a very small number close to zero. In addition, the error is aggressively amplified by the numerator \((1.0+x)\), leading to a significant error. However, Table~\ref{sample_82} shows that the outputs of both the GSL function and the \textit{mpmath} version are nearly identical as the input is approaching 1. 
This suggests that high-precision programs may not always be sufficient for detecting all potential errors.

\textbf{High-precision programs exhibit significantly slower execution time compared to original GSL programs.}  Table~\ref{motivation} demonstrates the results, across key statistical metrics. GSL functions yield results within a narrow range. GSL functions achieve consistent performance, with execution time in the range of 0.0003 to 0.0010 seconds and an average of 0.0005 seconds. 
In contrast, high-precision implementations using \textit{mpmath} exhibit much broader variability in execution times, ranging from 0.0094 seconds to over 60 seconds, with an average runtime of 2.6358 seconds. This equates to approximately 5,272 times the runtime of their GSL counterparts. 
This discrepancy further emphasizes the challenges associated with high-precision computations and the need for robust error detection in these implementations.

\begin{table}[tb]
\centering
\footnotesize
\caption{Execution time (second) of original GSL functions and \textit{mpmath} programs.}
\label{motivation}

\begin{tabular}{lcccr}
\toprule
  & \textbf{Min} & \textbf{Max} & \textbf{Mean} & \textbf{Median} \\
\midrule
\textbf{GSL}               & 0.0003 & 0.0010 & 0.0005 & 0.0005 \\
\textbf{Mpmath}   & 0.0094 & 60.0495 & 2.6358& 0.1978\\

\bottomrule
\end{tabular}

\end{table}

\findingboxx{According to our preliminary study, we find high-precision programs, though beneficial for error detection, pose several drawbacks, including the complexity of implementation, existence of potential faults, and long execution time. We believe these drawbacks can have a significant negative impact on software tasks, such as numerical software testing and program bug detection. Firstly, the complexity of rewriting high-precision programs to obtain oracles significantly raises the threshold for testing. Secondly, due to the faults, developers may overlook regions with significant errors. Thirdly, the long execution time of high-precision programs makes it impractical to use them for testing complex numerical programs.
}

%% file: definition.tex
\section{Problem Definition}
\label{definition}


\noindent\textbf{Novelty Analysis of DELA:} Prior research \cite{zou2019detecting} finds that a significant error in the result is often caused by one or more substantial amplifications from atomic operations, such as +, -, and \(sin(x)\). To address this issue, Zou et al. proposed ATOMU, an approach for detecting inputs that could lead to large condition numbers for atomic operations. However, the inputs may not lead to significant errors necessarily due to error masked \cite{bao2013fly} during program execution, such as in cases where a large number is added to a small number. Therefore, ATOMU could not be used to determine whether an input would lead to significant errors in a program. To the best of our knowledge, no existing approaches directly detect significant errors induced by large condition numbers in atomic operations. Recent efforts  \cite{wang2022detecting, zhang2023eiffel, zhang2024hierarchical} have developed methods to identify inputs likely to cause significant errors. However, these tools rely on high-precision programs to obtain oracles for error detection. 
The reliance on high-precision programs emphasizes the absence of effective tools for directly assessing numerical errors due to large condition numbers. 
Notably, recent works~\cite{wang2022detecting, zhang2023eiffel, zhang2024hierarchical} continue to use high-precision programs to assess whether a specific input may lead to a substantial error when using ATOMU as a baseline.

\noindent\textbf{Problem Definition:} Given a floating-point program and an input, \textbf{the objective} is to determine whether the input could cause significant errors due to large condition numbers in atomic operations. Prior study~\cite{zou2019detecting} lists common atomic operations and provides their corresponding condition numbers. Table~\ref{condition} selects the cases which could lead to significant condition numbers. Such ill-conditioned problems can lead to more troublesome numerical bugs compared to other issues, as they are less likely to be detected by standard compiler checks. 
To address this challenge, we propose DELA, a novel approach designed to detect these numerical errors.

\begin{table}[tb]
\centering
\footnotesize
\caption{Condition numbers and dangerous regions of atomic operations.}
\label{condition}

\begin{tabular}{lcr}
\toprule
 \textbf{Operation (\(op\))} & \textbf{Condition Number (\(C_{op}\))} & \textbf{Dangerous Region}\\
\midrule
\(op(x,y)=x+y\)   &  \(C_{+,x}(x,y)=\left|\frac{x}{x+y}\right|, C_{+,y}(x,y)=\left|\frac{y}{x+y}\right|\) & \(x\approx-y\)\\
  \(op(x,y)=x-y\)   & \(C_{-,x}(x,y)=\left|\frac{x}{x-y}\right|,C_{-,y}(x,y)=\left|-\frac{y}{x-y}\right|\)  & \(x\approx y\) \\
\(op(x)=\sin(x)\) &  \(C_{\sin}(x)=|x\cdot\cot(x)|\) & \(x\to n\pi,n\in\mathbb{Z}\) \\
\(op(x)=\cos(x)\) & \(C_{\cos}(x)=|x\cdot\tan(x)|\) & \(x\to n\pi+\frac\pi2,n\in\mathbb{Z}\) \\
\(op(x)=\tan(x)\) & \(C_{\tan}(x)=\left|\frac x{\sin(x)\cos(x)}\right|\) & \(x\to\frac{n\pi}2,n\in\mathbb{Z}\) \\
\(op(x)=\arcsin(x)\) & \(C_{\arcsin}(x)=\left|\frac{x}{\sqrt{1-x^2}\cdot\arcsin(x)}\right|\) & \(x\to-1^+,x\to1^-\) \\
\(op(x)=\arccos(x)\) & \(C_{\arccos}(x)=\left|-\frac{x}{\sqrt{1-x^2}\cdot\arccos(x)}\right|\) & \(x\to-1^+,x\to1^-\) \\
\(op(x)=\sinh(x)\)  & \(C_{\sinh}(x)=|x\cdot\coth(x)|\) & \(x\to\pm\infty \) \\
\(op(x)=\cosh(x)\) & \(C_{\cosh}(x)=|x\cdot\tanh(x)|\) &   \(x\to\pm\infty \) \\
\(op(x)=\exp(x)\) & \(C_{\exp}(x)=|x|\) &   \(x\to\pm\infty \)\\
\(op(x)=\log(x)\) & \(C_{\log}(x)=\left|\frac1{\log x}\right|\) & \(x \to 1\) \\
\(op(x)=\log_{10}(x)\) & \(C_{\log10}(x)=\left|\frac1{\log x}\right|\) & \(x \to 1\) \\
\(op(x,y)=x^y\)  & \(C_{pow,x}(x,y)=|y|,C_{pow,y}(x,y)=|y\log(x)|\) & \(x\to0^+,y\to\pm\infty \) \\

\bottomrule
\end{tabular}

\end{table}

%% file: methodology.tex
\section{Methodology}
\label{method}

\begin{figure*}[htbp]
\centerline{\includegraphics[width=1\textwidth]{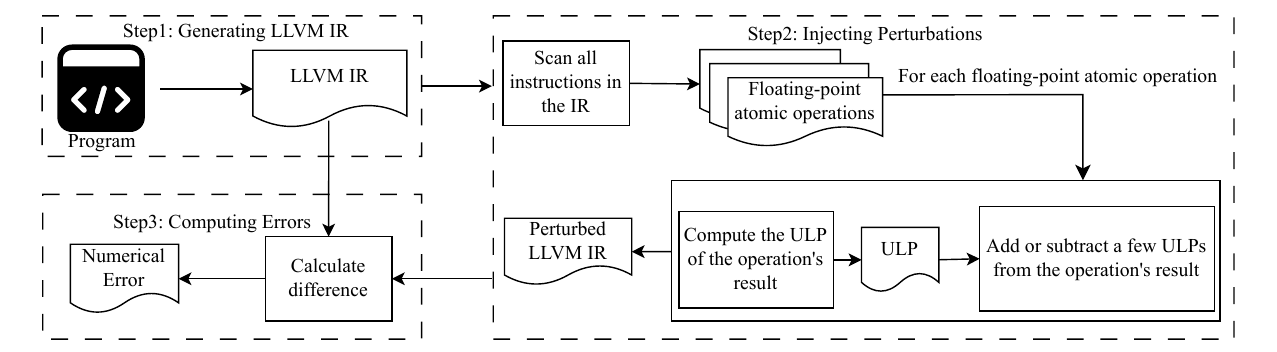}}
\caption{An overview of DELA.}
\label{framework}
\end{figure*}

In this section, we provide a detailed explanation of our approach, named DELA. First, we generate LLVM Intermediate Representation (IR) \cite{lattner2004llvm} of the target program. Next, we inject small perturbations into the outputs of atomic operations. Finally, we calculate the differences between the results of the original program and those of the perturbed version. To clarify this process, we also provide a concrete example in Section~\ref{motivate} demonstrating the workflow of our approach. An overview of our approach is shown in Figure~\ref{framework}. 

\subsection{Generating LLVM IR}
 This initial step involves compiling floating-point programs into LLVM IR, enabling us to leverage LLVM's optimizer to introduce perturbations directly into the program, rather than requiring specialized analysis for each individual case. Moreover, many programming languages, such as C, C++, Go, and Python, could generate LLVM IR based on the source code \cite{sarda2015llvm}. As a result, DELA has the potential to serve as a general-purpose tool applicable to a wide range of programs.


\subsection{Injecting Perturbations}
The perturbation injection is implemented as an LLVM pass in DELA. First, DELA scans the LLVM IR instruction by instruction. Afterward, our approach injects a function call whenever it encounters floating-point atomic operations. Next, the function calculates the ULP of the atomic operation's result and returns the modified result by adding or subtracting a few ULPs. Consequently, DELA introduces perturbations into the program smoothly, simulating the rounding errors inherent in the original program.

\subsection{Computing Errors}
After injection, the instrumented LLVM IR is compiled into a library. Any client program can use this library to obtain the error by calculating the difference between the original program and the perturbed version. DELA incurs minimal overhead compared to high-precision programs because it only performs two additional operations, i.e., ULP calculation and the modified result computation. This efficient approach ensures that DELA remains a practical solution without excessive computational cost.

\subsection{Illustrative example}

We use the example in Section~\ref{motivate} to demonstrate DELA's workflow. 
Table~\ref{method_} records the intermediate results of the function's execution taking 0.9999999999999809 as the input. We record the values of variables from both the original and the perturbed processes.


\begin{table}[tb]
\centering
\footnotesize
\caption{Main steps of \textit{gsl\_sf\_legendre\_Q0\_e} function by using our approach.}
\label{method_}

\begin{tabular}{lccr}
\toprule
\textbf{Step}  & \textbf{Original Process} & \textbf{Perturbed Process} & \textbf{ULP Difference}  \\
\midrule
temp1 = 1.0 - x & 1.\textbf{9095836023552692}e-14 & 1.\textbf{8984813721090177}e-14 & 3.52E+13 \\

 temp2 = (1.0 + x) / temp1  & 10\textbf{4734875055126.81}&  10\textbf{5347359704572.0} & 3.92E+13 \\

result = 0.5 * log(temp2)  & 16.14\textbf{1226636711792} & 16.14\textbf{414209686719} & 8.21E+11 \\

\bottomrule
\end{tabular}

\end{table}

First, \(x\) is perturbed during the scanning process, the value of \textit{temp1} is changed significantly because of the large condition number. Specifically, the condition number of \(1 - x\) respect to \(x\) is \(\left|-\frac{x}{1-x}\right|\) whose value is 5.2356e+13. Then the relative error of \(1 - x\) corresponds to the perturbation in \(x\) multiplied by 5.2356e+13, resulting in a significant error. Moreover, the error is significantly amplified during the division step. Consequently, the final result error remains excessively large.
Figure~\ref{subfigure2} shows the illustrative result including a set of inputs near 1. Notably, the errors produced by DELA form a perfect mountain peak in the figure, demonstrating DELA's capability to detect bugs within a specific region. Since our approach is built on LLVM IR and packaged into an easy-to-use library, it can completely resolve the challenge associated with implementation, i.e., the first challenge. 

%% file: evaluation.tex
\section{Evaluation}
\label{eva}
In Section~\ref{method}, we perform empirical evaluations to assess DELA’s effectiveness using verified significant errors across multiple datasets. Then we classify the faults of using high-precision to detect numerical errors into three categories and gauge whether we could overcome them. Furthermore, a significant challenge posed by high-precision programs is their prolonged execution times, which can hinder effective software engineering tasks, e.g., bug detection and program fixing. 




\subsection{Experiment Setup}

\subsubsection{Dataset}

We base our evaluation on datasets from the open-source packages of ATOMU \cite{zou2019detecting} and HSED \cite{zhang2024hierarchical}, both designed to find inputs that lead to significant errors. First, we need a set of verified significant errors to assess whether our approach can successfully detect them. ATOMU considers a relative error greater than 0.1\% as significant and obtains some inputs. However, some of these inputs are excessively large, which are not common \cite{zhang2024hierarchical}. Therefore, we select 49 out of these inputs, with values constrained within the half-precision floating-point range (i.e., [-65,504.0, -65,504.0]), as our evaluation data. For HSED, we also select the cases whose relative errors are more than 0.1\%, resulting in two cases. 

Existing research tends to focus on small numerical programs, such as GSL functions, due to the excessive time required for high-precision computations on more complex systems. 
To evaluate the potential applicability of DELA for more complex programs, we use a linear system-solving program as a complex test case, a core area of numerical linear algebra. When the condition number of a matrix is large, solving the corresponding linear system accurately becomes challenging \cite{ford2014numerical}. We hypothesize that the inaccuracy can be attributed to large atomic condition numbers and that DELA can detect such errors. Thus, we generate a series of matrices with varying condition numbers and employ DELA to identify significant errors in solving these linear systems.

\subsubsection{Evaluation Process}
We conduct our evaluation on three datasets to verify whether DELA can detect significant errors. 
\begin{itemize}
    \item \textbf{(1) ATOMU.} We first select 49 cases out of the ATOMU's results, which contain significant errors. Next, for each of these cases, we extend 1,000 points to the left and right, using intervals of ten times the 64-bit ULP. For each of these extended points, we compute the error using both our approach and a high-precision program.
    Afterwards, to further evaluate the reliability of DELA, we assess whether the errors calculated by high-precision programs for points near these inputs are strongly correlated with those produced by our approach. Particularly, we calculate the Spearman and Pearson correlation values between the two sets of errors.

    \item \textbf{(2) HSED.} This dataset contains only two cases that possess significant errors. We apply a process similar to that used for ATOMU, extending points and calculating the correlation values. However, the interval used is one 32-bit ULP, as the HSED programs are relatively simple and less sensitive to changes in input. 
    
    \item \textbf{(3) 3,000-dimensional linear system-solving program.} 
    To ensure that the linear system-solving program's runtime is neither too long nor too short, we chose 3,000-dimensional matrices for the experimental data. We generate a dataset of 1,000 matrices, some of which are randomly modified to be nearly singular to exhibit huge matrix condition numbers and numerical instability. We use the \textit{lu\_solve} function of GSL to implement normal-precision linear system solving. For the high-precision program, we employ the \textit{Rgesv} function of \textit{mplapack}\footnote{https://github.com/nakatamaho/mplapack} with 128-bit floating point numbers, which also performs the LU Decomposition to solve linear systems. We consider the three largest errors calculated by \textit{Rgesv} as significant errors and investigate whether the errors for these three cases, derived from our method, are also substantial.
\end{itemize}

In our preliminary and normal experiments, some errors calculated using DELA differ from those obtained through high-precision programs. We discover that these discrepancies arise from faults when using high-precision programs to detect numerical errors. We classify these faults into different categories and demonstrate that DELA can overcome these faults.

To evaluate whether DELA can reduce execution time compared to traditional high-precision methods, we involve the following steps. First, we collect a set of 51 inputs from the ATOMU and HSED datasets.
Next, we expand our analysis by calculating the errors for an additional 2,000 surrounding values for each of the selected inputs. This expansion is crucial for understanding how DELA's performance scales with varying input parameters. We execute these calculations using both DELA and the corresponding high-precision programs to create a comprehensive dataset for comparison.
Afterward, we record the total execution time for each case. 
In addition to the ATOMU and HSED datasets, we extend our evaluation to include complex numerical scenarios by examining the execution times of 3,000-dimensional linear system-solving programs when employing DELA. For these cases, we compare the execution time of DELA with that of high-precision counterparts.

\subsubsection{Evaluation Metric}
We utilize \(Err_{ulp}\) (c.f. Section~\ref{bac}) to quantify errors. For correlation analysis, we calculate both the Pearson and Spearman correlation coefficients to assess the association between errors detected by DELA and high-precision programs. The Pearson correlation coefficient measures the strength of the linear association between two variables \cite{sedgwick2012pearson}. The Spearman correlation coefficient, on the other hand, is defined as the Pearson correlation coefficient between the rank variables \cite{myers2013research}, assessing monotonic relationships.

\subsubsection{Software and Hardware Environment}
Our primary experimental environment is a Docker container running Ubuntu 24.04 on a laptop equipped with an AMD Ryzen 7 6800H @ 3.20 GHz CPU and 16GB RAM. However, for solving systems of linear equations, we employ a separate Docker container with Ubuntu 22.04 on a system with a 13th Gen Intel(R) Core(TM) i9-13900K CPU and 128GB RAM to accommodate higher computational demands.

\subsection{Detected Results}

\noindent\textbf{DELA successfully detects all significant bugs from the ATOMU and HSED datasets.} Table~\ref{error_and_correlation} presents the results of errors calculated by DELA and high-precision programs. Note from Table~\ref{error_and_correlation}, DELA can detect all the 51 significant errors, except for the ATOMU\_31, ATOMU\_32, and ATOMU\_33 cases. These three missed errors were actually attributed to the fault of high-precision programs in detecting numerical errors, which we will discuss in the next subsection. The errors calculated by DELA tend to be larger than those of high-precision programs because the perturbations we inject are usually larger than typical rounding errors in atomic operations. Additionally, values of some inputs in Table~\ref{error_and_correlation} are so close that they cannot be distinguished with four decimal places in base 10.

\begin{table}[]
\centering
\caption{Errors calculated by DELA and high-precision programs. The correlation between errors of points near the input computed by DELA and those from high-precision programs. }
\label{error_and_correlation}
\resizebox{0.8\textwidth}{!}{
\begin{tabular}{lccccccr}
\toprule
\multicolumn{1}{c}{\multirow{2}{*}{\textbf{Index}}} & \multicolumn{1}{c}{\multirow{2}{*}{\textbf{\begin{tabular}[c]{@{}c@{}}Input\\ Value\end{tabular}}}} & \multicolumn{1}{c}{\multirow{2}{*}{\textbf{\begin{tabular}[c]{@{}c@{}}Error\\ of DELA\end{tabular}}}} & \multicolumn{1}{c}{\multirow{2}{*}{\textbf{\begin{tabular}[c]{@{}c@{}}Error of high-\\ precision program\end{tabular}}}} & \multicolumn{1}{c}{\multirow{2}{*}{\textbf{\begin{tabular}[c]{@{}c@{}}Pearson \\ Correlation\end{tabular}}}} & \multicolumn{1}{c}{\multirow{2}{*}{\textbf{P-value}}} & \multicolumn{1}{c}{\multirow{2}{*}{\textbf{\begin{tabular}[c]{@{}c@{}}Spearman \\ Correlation\end{tabular}}}} & \multicolumn{1}{c}{\multirow{2}{*}{\textbf{P-value}}} \\
\multicolumn{1}{c}{}                                & \multicolumn{1}{c}{}                                                                                & \multicolumn{1}{c}{}                                                                                  & \multicolumn{1}{c}{}                                                                                                     & \multicolumn{1}{c}{}                                                                                         & \multicolumn{1}{c}{}                                  & \multicolumn{1}{c}{}                                                                                          & \multicolumn{1}{c}{}                                                  \\ \midrule
ATOMU\_1          & -1.1737                               & 8.74E+18                              & 7.28E+14                              & 0.7852                                & 0.00E+00                              & 0.5118                                & 4.93E-134                             \\
ATOMU\_2          & -1.1737                               & 1.02E+16                              & 5.90E+14                              & 0.7804                                & 0.00E+00                              & 0.5085                                & 4.81E-132                             \\
ATOMU\_3          & -1.0188                               & 2.25E+15                              & 1.83E+13                              & 0.7135                                & 0.00E-02                              & 0.5331                                & 2.43E-147                             \\
ATOMU\_4          & -2.2944                               & 4.50E+15                              & 1.78E+15                              & 0.9609                                & 0.00E+00                              & 0.5066                                & 6.27E-131                             \\
ATOMU\_5          & -29171.9364                           & 3.94E+15                              & 3.11E+14                              & 0.8404                                & 0.00E+00                              & 0.5889                                & 4.00E-187                             \\
ATOMU\_6          & -1.0188                               & 1.00E+15                              & 6.37E+13                              & 0.7815                                & 0.00E+00                              & 0.5343                                & 4.27E-148                             \\
ATOMU\_7          & -2.2944                               & 1.58E+15                              & 3.11E+13                              & 0.4936                                & 2.31E-123                             & 0.5113                                & 1.06E-133                             \\
ATOMU\_8          & -2.2944                               & 2.36E+15                              & 1.08E+14                              & 0.4465                                & 1.24E-98                              & 0.5092                                & 1.74E-132                             \\
ATOMU\_9          & 5.5201                                & 1.51E+15                              & 2.98E+13                              & 0.9017                                & 0.00E+00                              & 0.5979                                & 3.06E-194                             \\
ATOMU\_10         & -2.4048                               & 8.73E+18                              & 3.16E+15                              & 0.9995                                & 0.00E+00                              & 0.5607                                & 4.64E-166                             \\
ATOMU\_11         & 3.8317                                & 1.24E+16                              & 1.76E+13                              & 0.4983                                & 4.45E-126                             & 0.3111                                & 3.77E-46                              \\
ATOMU\_12         & 3.9577                                & 4.22E+14                              & 1.53E+14                              & 0.9944                                & 0.00E+00                              & 0.6675                                & 2.57E-258                             \\
ATOMU\_13         & 5.4297                                & 1.53E+15                              & 1.50E+13                              & 0.7904                                & 0.00E+00                              & 0.6254                                & 1.14E-217                             \\
ATOMU\_14         & 2.1971                                & 5.38E+15                              & 3.49E+14                              & 0.9415                                & 0.00E+00                              & 0.6858                                & 4.17E-278                             \\
ATOMU\_15         & -7.7253                               & 3.05E+15                              & 1.58E+13                              & 0.9240                                & 0.00E+00                              & 0.6645                                & 3.27E-255                             \\
ATOMU\_16         & 9.0950                                & 4.50E+15                              & 3.85E+13                              & 0.9023                                & 0.00E+00                              & 0.5707                                & 2.61E-173                             \\
ATOMU\_17         & 2.7984                                & 1.21E+15                              & 1.73E+14                              & 0.6818                                & 1.01E-273                             & 0.5478                                & 4.12E-157                             \\
ATOMU\_18         & 3.9595                                & 3.38E+15                              & 7.66E+14                              & 0.9942                                & 0.00E+00                              & 0.6581                                & 1.14E-248                             \\
ATOMU\_19         & 3.1416                                & 4.50E+15                              & 4.55E+15                              & 0.8265                                & 0.00E+00                              & 0.4650                                & 7.06E-108                             \\
ATOMU\_20         & 12.5952                               & 1.41E+16                              & 5.54E+15                              & 0.8503                                & 0.00E+00                              & 0.5998                                & 8.90E-196                             \\
ATOMU\_21         & -0.3725                               & 1.97E+15                              & 1.68E+14                              & 0.9237                                & 0.00E+00                              & 0.6702                                & 3.77E-261                             \\
ATOMU\_22         & -1.3472                               & 1.55E+16                              & 1.55E+16                              & \textbf{0.0265}                       & \textbf{2.35E-01}                     & 0.4709                                & 6.00E-111                             \\
ATOMU\_23         & -0.3725                               & 2.33E+15                              & 2.31E+14                              & 0.9520                                & 0.00E+00                              & 0.6553                                & 7.53E-246                             \\
ATOMU\_24         & 0.3725                                & 5.63E+14                              & 9.15E+13                              & 0.9360                                & 0.00E+00                              & 0.7316                                & 0.00E+00                              \\
ATOMU\_25         & 0.3725                                & 1.01E+15                              & 1.06E+15                              & 0.9493                                & 0.00E+00                              & 0.7166                                & 0.00E-02                              \\
ATOMU\_26         & 0.5238                                & 1.13E+15                              & 3.81E+14                              & 0.9733                                & 0.00E+00                              & 0.3518                                & 2.35E-59                              \\
ATOMU\_27         & 0.6165                                & 1.01E+16                              & 3.37E+15                              & 0.9295                                & 0.00E+00                              & 0.5037                                & 3.40E-129                             \\
ATOMU\_28         & 9.5256                                & 2.81E+14                              & 4.97E+13                              & 0.9927                                & 0.00E+00                              & 0.6605                                & 3.85E-251                             \\
ATOMU\_29         & -2.4570                               & 8.78E+18                              & 3.50E+15                              & 0.4031                                & 4.83E-79                              & 0.7285                                & 0.00E+00                              \\
ATOMU\_30         & -4.9915                               & 8.81E+18                              & 2.63E+13                              & 0.9782                                & 0.00E+00                              & 0.5328                                & 3.72E-147                             \\
ATOMU\_31         & \textbf{-6.15E-42}                    & \textbf{0.00E+00}                     & \textbf{4.18E+15}                     & 0.0000                                & 0.00E+00                              & 0.0000                                & 0.00E+00                              \\
ATOMU\_32         & \textbf{2.32E-144}                    & \textbf{0.00E+00}                     & \textbf{4.69E+179}                    & 0.0000                                & 0.00E+00                              & 0.0000                                & 0.00E+00                              \\
ATOMU\_33         & \textbf{7.10E-42}                     & \textbf{0.00E+00}                     & \textbf{1.72E+15}                     & 0.0000                                & 0.00E+00                              & 0.0000                                & 0.00E+00                              \\
ATOMU\_34         & -0.5774                               & 8.73E+18                              & 3.94E+14                              & 0.9602                                & 0.00E+00                              & 0.5067                                & 5.95E-131                             \\
ATOMU\_35         & -0.7746                               & 2.76E+15                              & 6.07E+14                              & 0.9498                                & 0.00E+00                              & 0.5104                                & 3.30E-133                             \\
ATOMU\_36         & 0.8336                                & 1.35E+16                              & 9.11E+15                              & 0.9908                                & 0.00E+00                              & 0.6188                                & 8.44E-212                             \\
ATOMU\_37         & -42.0000                              & 2.50E+15                              & 3.24E+14                              & 0.8269                                & 0.00E+00                              & 0.6197                                & 1.41E-212                             \\
ATOMU\_38         & -5.6672                               & 1.23E+16                              & 7.85E+15                              & 0.9872                                & 0.00E+00                              & 0.6065                                & 2.59E-201                             \\
ATOMU\_39         & -31640.0000                           & 1.28E+15                              & 1.28E+14                              & 0.9660                                & 0.00E+00                              & 0.7017                                & 1.08E-296                             \\
ATOMU\_40         & -1.0000                               & 3.58E+14                              & 7.80E+13                              & 0.9786                                & 0.00E+00                              & 0.1387                                & 4.66E-10                              \\
ATOMU\_41         & 0.8814                                & 4.38E+18                              & 1.19E+15                              & 0.9073                                & 0.00E+00                              & 0.5876                                & 4.13E-186                             \\
ATOMU\_42         & -10.0000                              & 4.36E+18                              & 6.44E+12                              & 0.1043                                & 2.96E-06                              & 0.7280                                & 0.00E+00                              \\
ATOMU\_43         & -170.0000                             & 2.82E+15                              & 1.17E+13                              & 0.9942                                & 0.00E+00                              & 0.7501                                & 1.32E-181                             \\
ATOMU\_44         & -170.0000                             & 2.82E+15                              & 1.17E+13                              & 0.9942                                & 0.00E+00                              & 0.7501                                & 1.34E-181                             \\
ATOMU\_45         & -10.0000                              & 3.04E+15                              & 6.85E+13                              & 0.8293                                & 0.00E+00                              & 0.7252                                & 0.00E+00                              \\
ATOMU\_46         & -10.0000                              & 5.69E+15                              & 3.08E+14                              & 0.9920                                & 0.00E+00                              & 0.7171                                & 0.00E-02                              \\
ATOMU\_47         & -170.0000                             & 1.75E+15                              & 1.07E+13                              & 0.9820                                & 0.00E+00                              & 0.7476                                & 9.25E-180                             \\
ATOMU\_48         & -170.0000                             & 1.75E+15                              & 1.07E+13                              & 0.9820                                & 0.00E+00                              & 0.7476                                & 9.25E-180                             \\
ATOMU\_49         & -170.0000                             & 2.82E+15                              & 1.17E+13                              & 0.9942                                & 0.00E+00                              & 0.7501                                & 1.34E-181                             \\
HSED\_1           & 6.2832                                & 5.70E+13                              & 2.85E+13                              & 0.9993                                & 0.00E+00                              & 0.8373                                & 0.00E+00                              \\
HSED\_2           & 6.2832                                & 4.77E+13                              & 2.38E+13                              & 0.9991                                & 0.00E+00                              & 0.8171                                & 0.00E+00                                                                        
\\ \bottomrule   
\end{tabular}
}
\end{table}

\begin{wrapfigure}{r}{0.6\textwidth}
    \centering
    \includegraphics[width=0.58\textwidth]{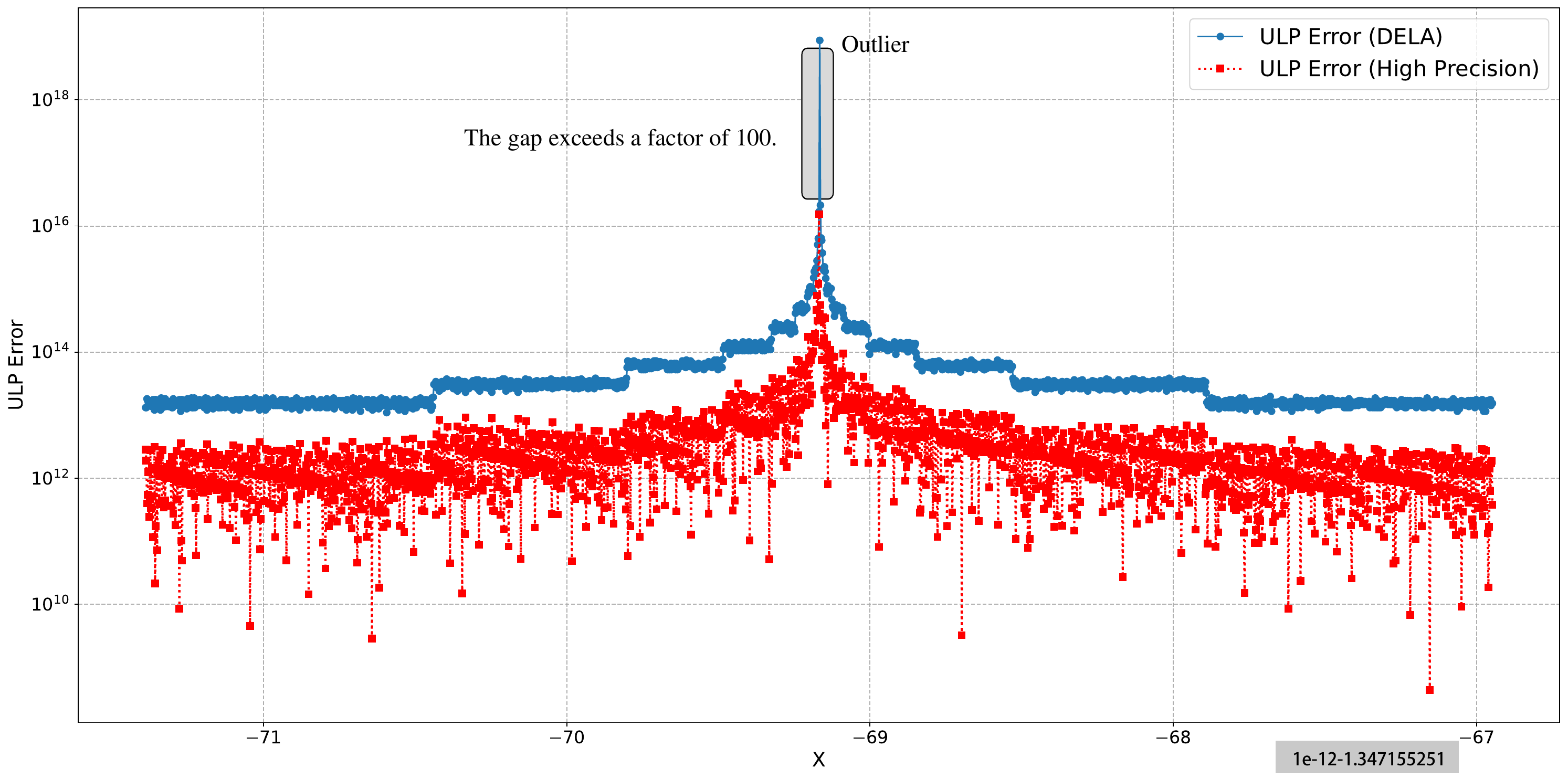}
    \caption{An outlier of DELA's errors.}
    \label{outlier}
\end{wrapfigure}

\textbf{DELA exhibits a strong correlation with high-precision programs' results.}
Table~\ref{error_and_correlation} shows that the average Pearson correlation score is 0.7962, while the average Spearman correlation score is 0.5694. 
These scores indicate a strong relationship between DELA’s results and the high-precision outputs \cite{kirbas2014effect}. Such a result suggests that DELA’s error assessments are closely aligned with those obtained from high-precision computations. 
We set the significance level as 0.05. However, the ATOMU\_22's Pearson correlation is only 0.0265 and the corresponding p-value is 2.35E-01, due to an outlier of errors calculated by DELA. As shown in Figure~\ref{outlier}, it exists near the input -1.3472. This outlier does not undermine our approach. On the contrary, one will easily observe this extra large error when they use DELA. Upon removing this outlier, the Pearson correlation score improves to 0.9101.



\begin{wrapfigure}{l}{0.6\textwidth}
    \centering
    \includegraphics[width=0.58\textwidth]{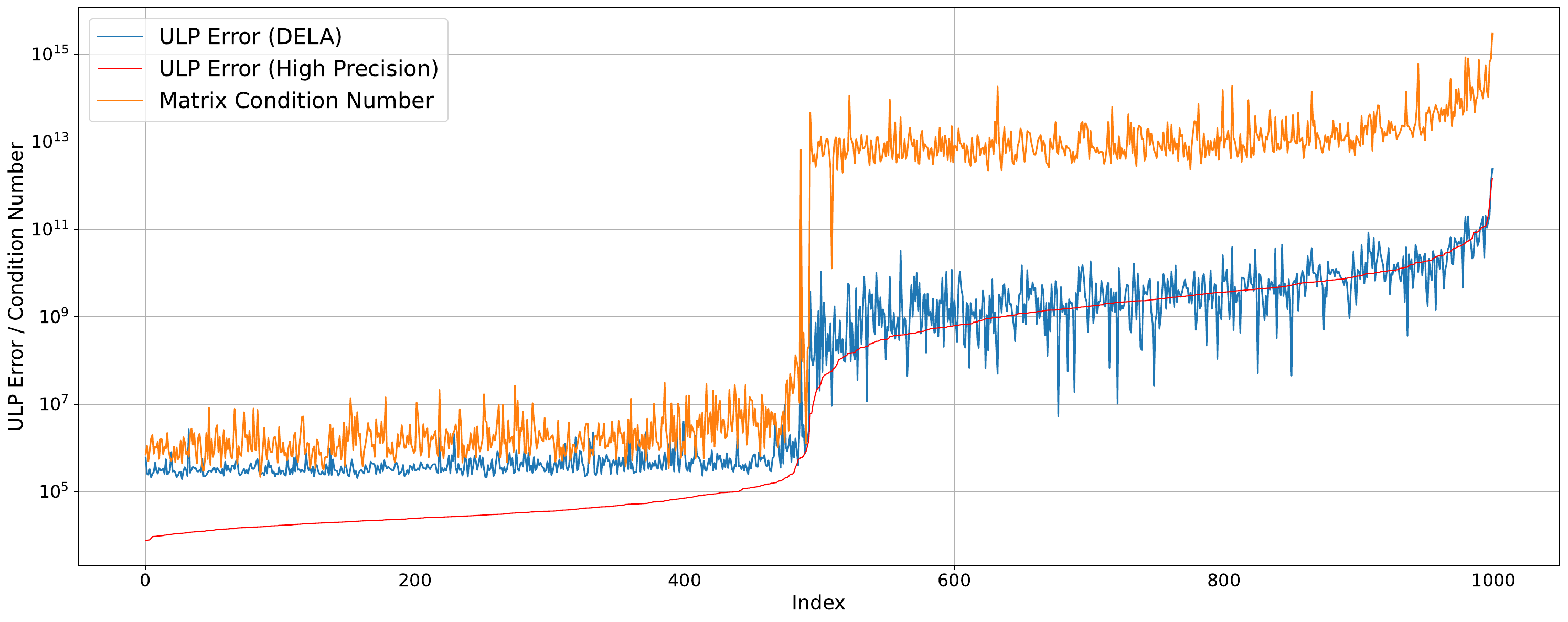}
    \caption{Errors calculated by DELA and high-precision programs and values of matrix condition number.}
    \label{correlation_matrix}
\end{wrapfigure}

\textbf{DELA demonstrates exceptional effectiveness in complex numerical programs.}
We obtain the oracles of the 1,000 cases using high-precision programs and select the three cases with the largest errors as ones with significant errors. DELA then computes the errors for all 1,000 cases, the rankings of the three cases are 1, 2, and 3, respectively. Therefore, in the context of the 3,000-dimensional linear system-solving program, DELA identifies all the significant errors. This remarkable performance highlights DELA’s robustness when confronted with complex scenarios.
Additionally, the Pearson correlation and Spearman correlation scores between the errors computed by DELA and the high-precision programs are 0.9763 and 0.8993, respectively, as shown in Figure~\ref{correlation_matrix}. Notably, the matrix condition numbers also illustrate a high correlation with errors computed by DELA, which is consistent with the theory of numerical linear algebra \cite{ford2014numerical}. As shown in Figure~\ref{gsl_linalg_LU_decomp}, we find nearly all the large condition number cases are catastrophic cancellations. 

\findingboxx{DELA effectively identifies significant numerical errors in the ATOMU and HSED datasets, achieving a strong correlation with high-precision programs. This robust performance emphasizes DELA's reliability, particularly in complex numerical programs. These findings suggest DELA’s suitability as a practical alternative for error detection in both GSL and complex numerical applications.}


\subsection{Faults When Using High-Precision Programs to Detect Numerical Errors}
According to our preliminary experiments and the detected results, several errors obtained from DELA differ from those generated by high-precision programs. Given that these high-precision programs are developed by expert software engineers and scientists, the unexpected discrepancies raise important questions about their implementation accuracy and integrity. We take the function \textit{gsl\_sf\_legendre\_Q0\_e} as an example in Section \ref{motivate}, indicating that using high-precision oracle for testing can miss one significant error. In this subsection, we classify these faults into three categories and demonstrate DELA can overcome these faults.


\textbf{Our experiments reveal three categories of faults when using high-precision programs to detect numerical errors.}

\begin{figure}[htbp]
    \centering
    \begin{minipage}{0.45\textwidth} 
        \centering
        \includegraphics[width=\textwidth]{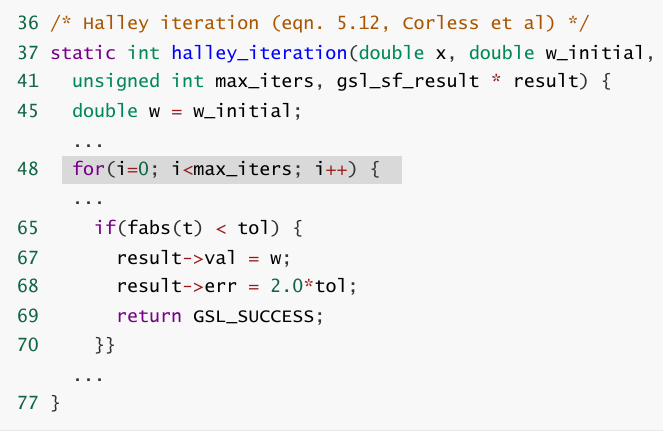}
        \subcaption{Code snippet of GSL function.}\label{1subfigure1}
    \end{minipage}
    \hfill 
    \begin{minipage}{0.45\textwidth} 
        \centering
        \includegraphics[width=\textwidth]{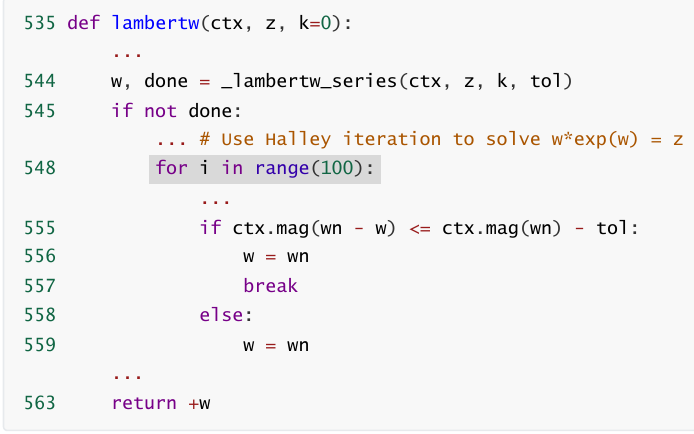}
        \subcaption{Code snippet of \textit{mpmath} function.}\label{1subfigure2}
    \end{minipage}
    \caption{Core code of the Halley iteration in the example.} 
    \label{func77_code}
\end{figure}

\textbf{Category 1: Different implementations in programs compared to their high-precision counterparts.} Poor performance of ATOMU\_31, ATOMU\_32, and ATOMU\_33 in Table~\ref{error_and_correlation} falls into this category of fault. For instance, the fault of ATOMU\_31 originates in the function \textit{gsl\_sf\_lambert\_W0\_e}, which applies Halley iteration when the input approaches 0. Meanwhile, Figure~\ref{func77_code} illustrates that \textit{mpmath} also relies on Halley iteration. However, the values of \textit{w} are 0.17966771 and -6.15166707e-42 for the two programs, respectively. Furthermore, the GSL function requires six iterations for this input, while the \textit{mpmath} version converges in only one iteration, resulting in a notable discrepancy. Therefore, we speculate that this issue stems from differences in implementation rather than an actual numerical bug in the GSL function. This discrepancy can potentially mislead developers and users, resulting in unproductive efforts in software engineering tasks. In contrast, DELA is grounded in the original program and can overcome this fault in different implementations.

 \textbf{Category 2: High-precision and low-precision programs exhibit the same numerical errors.} The function \textit{gsl\_sf\_legendre\_Q0\_e}, discussed in Section \ref{motivate}, can be classified into this category. For a given input, both the GSL function and its high-precision counterpart produce nearly identical results, even though a segment of numerical instability exists within the program (see Section \ref{motivate}). This similarity in outputs causes developers to overlook the bug, leaving critical vulnerabilities unaddressed. In contrast, DELA could find the numerical error.

\textbf{Category 3: Limited numerical range of low-precision programs.} In the case of the function \textit{gsl\_sf\_zeta\_e}, when the input is less than -170, an overflow error is triggered, leading to the termination of the current calculation and the return of error status. As noted by the developer in the source code, ``The actual zeta function may or may not overflow here. But we have no easy way to calculate it when the prefactor(s) overflow''. Consequently, as illustrated in Table~\ref{results_of_function104}, both the original and perturbed results yield a value of 0 when the input is less than -170, where we set the error status as 0. Therefore, the observed discrepancy can be attributed to the limited range of double-precision floating-point numbers rather than numerical bugs. Conversely, DELA would not detect this fake numerical error, because no numerical operations exist in the branch.

 \begin{table}[]
\centering
\caption{gsl\_sf\_zeta\_e's and the \textit{mpmath} version's results of points near -170.}
\label{results_of_function104}
\resizebox{0.85\textwidth}{!}{
\begin{tabular}{lccccr}
\toprule
\multicolumn{1}{c}{\multirow{2}{*}{\textbf{Input}}} & \multicolumn{1}{c}{\multirow{2}{*}{\textbf{\begin{tabular}[c]{@{}c@{}}Original \\ Result\end{tabular}}}} & \multicolumn{1}{c}{\multirow{2}{*}{\textbf{\begin{tabular}[c]{@{}c@{}}Perturbed \\ Result\end{tabular}}}} & \multicolumn{1}{c}{\multirow{2}{*}{\textbf{\begin{tabular}[c]{@{}c@{}}Ulp Error\\ of DELA\end{tabular}}}} & \multicolumn{1}{c}{\multirow{2}{*}{\textbf{\begin{tabular}[c]{@{}c@{}}mpmath \\ Result\end{tabular}}}} & \multicolumn{1}{c}{\multirow{2}{*}{\textbf{\begin{tabular}[c]{@{}c@{}}Ulp Error\\ of mpmath\end{tabular}}}} \\
\multicolumn{1}{c}{}                                & \multicolumn{1}{c}{}                                                                                     & \multicolumn{1}{c}{}                                                                                      & \multicolumn{1}{c}{}                                                                                      & \multicolumn{1}{c}{}                                                                                   & \multicolumn{1}{c}{}                                                                                        \\ \midrule
-170.00000000000014                                 & 0.0000E+00                                                                                               & 0.0000E+00                                                                                                & 0.0000E+00                                                                                                & 1.0515E+158                                                                                            & inf                                                                                                         \\
-170.0000000000001                                  & 0.0000E+00                                                                                               & 0.0000E+00                                                                                                & 0.0000E+00                                                                                                & 8.4117E+157                                                                                            & inf                                                                                                         \\
-170.00000000000009                                 & 0.0000E+00                                                                                               & 0.0000E+00                                                                                                & 0.0000E+00                                                                                                & 6.3088E+157                                                                                            & inf                                                                                                         \\
-170.00000000000006                                 & 0.0000E+00                                                                                               & 0.0000E+00                                                                                                & 0.0000E+00                                                                                                & 4.2059E+157                                                                                            & inf                                                                                                         \\
-170.00000000000003                                 & 0.0000E+00                                                                                               & 0.0000E+00                                                                                                & 0.0000E+00                                                                                                & 2.1029E+157                                                                                            & inf                                                                                                         \\
-170.0                                              & 0.0000E+00                                                                                               & 0.0000E+00                                                                                                & 0.0000E+00                                                                                                & 0.0000E+00                                                                                             & 0.0000E+00                                                                                                  \\
-169.99999999999997                                 & -2.1185E+157                                                                                             & 0.0000E+00                                                                                                & 6.9605E+18                                                                                                & -2.1029E+157                                                                                           & 5.1105E+13                                                                                                  \\
-169.99999999999994                                 & -4.2103E+157                                                                                             & -2.0976E+157                                                                                              & 4.5284E+15                                                                                                & -4.2059E+157                                                                                           & 7.3363E+12                                                                                                  \\
-169.99999999999991                                 & -6.3231E+157                                                                                             & -4.1894E+157                                                                                              & 2.8178E+15                                                                                                & -6.3088E+157                                                                                           & 1.1714E+13                                                                                                  \\
-169.9999999999999                                  & -8.4149E+157                                                                                             & -6.3022E+157                                                                                              & 1.7326E+15                                                                                                & -8.4117E+157                                                                                           & 2.6058E+12                                                                                                  \\
-169.99999999999986                                 & -1.0528E+158                                                                                             & -8.3940E+157                                                                                              & 1.7497E+15                                                                                                & -1.0515E+158                                                                                           & 1.0652E+13                       
                                                                                  
\\ \bottomrule
\end{tabular}
}
\end{table}

\findingboxx{We categorize the faults in high-precision programs into three distinct categories, all of which can be addressed by our approach. These faults may lead developers to incorrectly detect bugs or even overlook them, thereby hindering the completion of software engineering tasks.
}

\subsection{Execution Time Comparison}

\noindent\textbf{DELA can save excessive time compared to high-precision programs for both ATOMU and HSED datasets.} As shown in Table~\ref{time_table}, the cumulative execution time for all original GSL programs is recorded at approximately 0.08 seconds. After applying DELA's perturbation, the total execution time increases to about 0.34 seconds, indicating a growth of 3.3644 times. In contrast, the total execution time of high precision programs is about 265.38 seconds, reflecting a high rise of 3,376 times compared to the original programs. Consequently, the total runtime of DELA's perturbed programs is only 0.13\% of that of the high-precision programs. 
Such dramatic reductions underscore DELA’s efficiency in reducing execution time while still producing reliable results.

\noindent\textbf{DELA performs well for complex programs with long execution time.} As for the 1,000 3,000-dimensional linear system-solving programs, the total execution time of the original program is 0.6414 hours, rising to 0.9893 hours for the perturbed version. In contrast, the total runtime of the high-precision program is 73.6650 hours, making it 73.4599 times longer than the execution time for DELA’s perturbed version. This disparity illustrates DELA's capacity to effectively handle long runtimes typical of complex numerical simulations. For example, consider a numerical simulation program that typically requires a full day to complete \cite{he2020testing}. Using a high-precision program to evaluate a specific input for significant errors could extend the runtime to approximately 100 days. DELA, on the other hand, offers a practical solution, significantly reducing this time while still ensuring accurate error detection. 

\noindent\textbf{The analysis reveals a positive correlation between the number of perturbations executed by DELA and the corresponding execution time.} We observe Pearson and Spearman correlation scores of 0.9762 and 0.8866, respectively, indicating a high positive correlation. This means that as the number of perturbations increases, the execution time tends to rise correspondingly. 

\begin{table}[]
\centering
\caption{Runtime of 51 cases from ATOMU and HSED.}
\label{time_table}
\resizebox{0.8\textwidth}{!}{
\begin{tabular}{lcccccr}
\toprule
\multicolumn{1}{c}{\multirow{2}{*}{\textbf{Index}}} & \multicolumn{1}{c}{\multirow{2}{*}{\textbf{\begin{tabular}[c]{@{}c@{}}Original \\ GSL Time (ms)\end{tabular}}}} & \multicolumn{1}{c}{\multirow{2}{*}{\textbf{\begin{tabular}[c]{@{}c@{}}Perturbed \\ GSL Time (ms)\end{tabular}}}} & \multicolumn{1}{c}{\multirow{2}{*}{\textbf{\begin{tabular}[c]{@{}c@{}}High-precision \\ version time (ms)\end{tabular}}}} & \multicolumn{1}{c}{\multirow{2}{*}{\textbf{\begin{tabular}[c]{@{}c@{}}Numbers of\\ perturbation\end{tabular}}}} & \multicolumn{1}{c}{\multirow{2}{*}{\textbf{\begin{tabular}[c]{@{}c@{}}Perturbed GSL\\ /Original GSL\end{tabular}}}} & \multicolumn{1}{c}{\multirow{2}{*}{\textbf{\begin{tabular}[c]{@{}c@{}}High-precision\\ /Perturbed GSL\end{tabular}}}} \\
\multicolumn{1}{c}{}                                & \multicolumn{1}{c}{}                                                                                           & \multicolumn{1}{c}{}                                                                                            & \multicolumn{1}{c}{}                                                                                                     & \multicolumn{1}{c}{}                                                                                            & \multicolumn{1}{c}{}                                                                                                & \multicolumn{1}{c}{}                                                                                                  \\ \midrule
ATOMU\_1  & 3.30 & 12.77 & 345.66     & 3.8819E+05 & 386.83\% & 2,707.73\%     \\
ATOMU\_2  & 3.37 & 13.42 & 426.94     & 3.8819E+05 & 397.55\% & 3,182.19\%     \\
ATOMU\_3  & 3.33 & 13.02 & 385.28     & 4.0020E+05 & 390.58\% & 2,960.24\%     \\
ATOMU\_4  & 2.71 & 10.61 & 399.60     & 2.8814E+05 & 391.35\% & 3,767.57\%     \\
ATOMU\_5  & 1.90 & 7.40  & 519.24     & 2.2210E+05 & 389.58\% & 7,012.32\%     \\
ATOMU\_6  & 3.29 & 13.33 & 465.30     & 4.0020E+05 & 405.78\% & 3,490.60\%     \\
ATOMU\_7  & 2.61 & 9.77  & 481.69     & 2.8814E+05 & 373.90\% & 4,932.28\%     \\
ATOMU\_8  & 2.50 & 10.22 & 467.99     & 2.8814E+05 & 409.30\% & 4,580.36\%     \\
ATOMU\_9  & 2.92 & 11.98 & 37.17      & 3.7819E+05 & 410.81\% & 310.15\%       \\
ATOMU\_10 & 0.97 & 4.94  & 32.41      & 1.5808E+05 & 507.91\% & 656.38\%       \\
ATOMU\_11 & 0.94 & 4.90  & 34.26      & 1.4807E+05 & 519.46\% & 699.60\%       \\
ATOMU\_12 & 1.79 & 9.67  & 707.97     & 3.2216E+05 & 539.41\% & 7,318.11\%     \\
ATOMU\_13 & 2.65 & 12.17 & 1,974.50   & 3.8019E+05 & 459.01\% & 16,218.83\%    \\
ATOMU\_14 & 1.83 & 9.60  & 1,678.10   & 3.1416E+05 & 524.00\% & 17,477.48\%    \\
ATOMU\_15 & 0.27 & 0.76  & 47.25      & 1.4007E+04 & 284.85\% & 6,220.73\%     \\
ATOMU\_16 & 0.29 & 1.08  & 66.11      & 2.4011E+04 & 371.31\% & 6,109.18\%     \\
ATOMU\_17 & 1.70 & 7.45  & 45.48      & 2.0610E+05 & 437.47\% & 610.17\%       \\
ATOMU\_18 & 1.80 & 7.18  & 62.02      & 2.1611E+05 & 398.35\% & 863.33\%       \\
ATOMU\_19 & 1.25 & 6.08  & 7,130.43   & 1.8609E+05 & 484.94\% & 117,258.74\%   \\
ATOMU\_20 & 0.68 & 3.12  & 733.83     & 9.2046E+04 & 457.48\% & 23,507.69\%    \\
ATOMU\_21 & 1.18 & 6.51  & 57.47      & 1.9010E+05 & 553.40\% & 883.05\%       \\
ATOMU\_22 & 1.42 & 7.58  & 104.48     & 2.4212E+05 & 534.98\% & 1,379.04\%     \\
ATOMU\_23 & 1.24 & 6.10  & 76.96      & 1.9010E+05 & 494.10\% & 1,260.60\%     \\
ATOMU\_24 & 1.18 & 6.05  & 42.14      & 1.9010E+05 & 510.93\% & 696.19\%       \\
ATOMU\_25 & 1.18 & 6.12  & 78.91      & 1.9010E+05 & 517.15\% & 1,290.19\%     \\
ATOMU\_26 & 2.38 & 12.43 & 107.86     & 3.9420E+05 & 522.57\% & 867.59\%       \\
ATOMU\_27 & 1.06 & 5.30  & 39.44      & 1.7009E+05 & 498.97\% & 744.26\%       \\
ATOMU\_28 & 5.25 & 27.84 & 55.49      & 8.9044E+05 & 529.88\% & 199.33\%       \\
ATOMU\_29 & 0.57 & 2.86  & 52.76      & 8.8044E+04 & 503.12\% & 1,842.99\%     \\
ATOMU\_30 & 6.35 & 20.31 & 51.04      & 4.4222E+05 & 319.90\% & 251.30\%       \\
ATOMU\_31 & 0.32 & 0.40  & 78.03      & 1.1006E+05 & 123.04\% & 19,686.21\%    \\
ATOMU\_32 & 0.32 & 0.40  & 74.90      & 1.1006E+05 & 124.70\% & 18,901.29\%    \\
ATOMU\_33 & 0.32 & 0.40  & 72.37      & 1.1006E+05 & 123.26\% & 18,281.31\%    \\
ATOMU\_34 & 0.16 & 0.38  & 68.36      & 4.0020E+03 & 237.39\% & 18,221.76\%    \\
ATOMU\_35 & 0.20 & 0.77  & 67.08      & 2.0010E+04 & 380.07\% & 8,690.53\%     \\
ATOMU\_36 & 0.22 & 0.66  & 17,656.01  & 1.4007E+04 & 308.48\% & 2,661,542.57\% \\
ATOMU\_37 & 1.29 & 6.66  & 101.22     & 2.1011E+05 & 514.72\% & 1,519.79\%     \\
ATOMU\_38 & 1.30 & 6.53  & 107.59     & 2.1011E+05 & 500.52\% & 1,648.54\%     \\
ATOMU\_39 & 1.08 & 3.39  & 225,626.18 & 7.9192E+04 & 313.86\% & 6,650,930.94\% \\
ATOMU\_40 & 0.19 & 0.41  & 38.09      & 4.0020E+03 & 214.02\% & 9,403.73\%     \\
ATOMU\_41 & 0.39 & 1.34  & 32.91      & 3.6018E+04 & 346.18\% & 2,462.55\%     \\
ATOMU\_42 & 2.53 & 13.30 & 749.14     & 4.4218E+05 & 524.93\% & 5,632.38\%     \\
ATOMU\_43 & 0.44 & 1.50  & 440.47     & 3.3033E+04 & 339.03\% & 29,345.53\%    \\
ATOMU\_44 & 0.50 & 1.68  & 455.41     & 3.6036E+04 & 336.89\% & 27,178.72\%    \\
ATOMU\_45 & 2.66 & 14.00 & 881.63     & 4.5623E+05 & 527.02\% & 6,299.63\%     \\
ATOMU\_46 & 2.61 & 14.16 & 871.86     & 4.5623E+05 & 542.45\% & 6,156.88\%     \\
ATOMU\_47 & 0.49 & 1.70  & 429.52     & 3.6036E+04 & 349.30\% & 25,198.21\%    \\
ATOMU\_48 & 0.48 & 1.68  & 443.62     & 3.6036E+04 & 350.01\% & 26,362.80\%    \\
ATOMU\_49 & 0.48 & 1.68  & 449.36     & 3.6036E+04 & 348.08\% & 26,701.32\%    \\
HSED\_1   & 0.31 & 0.66  & 7.01       & 8.0040E+03 & 209.36\% & 1,069.02\%     \\
HSED\_2   & 0.34 & 0.66  & 18.16      & 8.0040E+03 & 193.12\% & 2,732.27\%         
\\  \bottomrule
\end{tabular}
}
\end{table}


\findingboxx{DELA significantly reduces execution time compared to high-precision programs while maintaining reliable error detection. 
The observed positive correlation between the number of perturbations and execution time indicates that DELA effectively balances execution speed with thorough error assessment. These findings suggest that integrating DELA into numerical analysis workflows can enhance efficiency and feasibility, especially in time-sensitive computational tasks.
}


%% file: discussion.tex
\section{Discussion}
\label{discussion}
In this section, we discuss the implementation of perturbations and numerical error detection and practical utility. 

\textbf{Implementation of perturbations.} In our experiments, one ULP subtraction serves as the primary perturbation, except one specific case. For the ATOMU\_11 and ATOMU\_17 in Table~\ref{error_and_correlation}, DELA could not detect the significant errors if we just subtract one ULP for all the results of atomic operations. This issue is due to the perturbation offset. Ideally, if the catastrophic cancellation is the root cause of significant errors and we subtract one ULP from one of the operands which is the result of another atomic operation, the original operation \(x-y\) transforms into \(x-(y-ULP)\). Consequently, the atomic operation would lead to large difference between the two results because of the huge condition number and the perturbation. However, if \(x\) is also reduced by one ULP, both results align, and the error is effectively masked. To address this, we apply a cyclic approach by subtracting and adding one, two, and three ULPs in succession.

\textbf{Numerical Error detection and practical utility.} One of the main motivations behind DELA’s design is to address the limitations of existing high-precision error detection methods. High-precision programs often require considerable execution time, rendering them impractical for larger-scale applications or real-time systems. DELA overcomes this by introducing perturbations directly at the atomic operation level, thereby emulating rounding errors without the computational cost of full high-precision computations. Our evaluation demonstrates the effectiveness of perturbation-based techniques in identifying numerical errors induced by large atomic condition numbers. This also provides a practical guideline for developers and researchers aiming to identify critical error-prone areas in their numerical programs. 


\begin{wrapfigure}{l}{0.5\textwidth}
    \centering
    \includegraphics[width=0.48\textwidth]{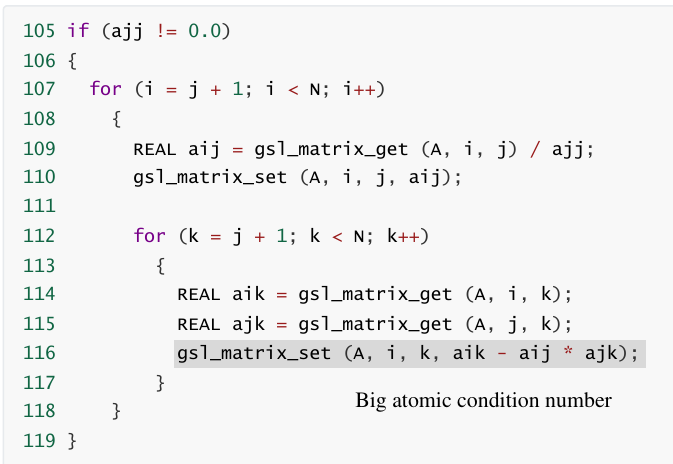}
    \caption{An example of big condition number in the 3,000-dimensional linear system-solving program.}
    \label{gsl_linalg_LU_decomp}
\end{wrapfigure}

%% file: threat.tex
\section{Threats to Validity}
\label{threat}
In this section, we discuss the threats to the validity of this work.

\noindent \textbf{External validity.} Our evaluation primarily focuses on C programs, which raises concerns about the generalizability of DELA across different programming languages. DELA may not perform as effectively with high-level languages such as Python and Java. However, it is important to note that DELA is built upon LLVM passes, enabling its application to any language that can be converted to LLVM Intermediate Representation (IR).

\noindent \textbf{Internal validity.} The choice of perturbation may introduce potential biases. In our experiments, we primarily use a single ULP subtraction as our perturbation method. However, we recognize that perturbation offsets may appear in two specific scenarios. To address this, we resolve it by using cyclic perturbation (c.f. Section \ref{discussion}). 
Nonetheless, reliance on a singular perturbation approach may result in DELA missing certain errors. This limitation suggests the need for further exploration of diverse perturbation strategies to enhance the approach's robustness.

\noindent \textbf{Construct validity.} Our approach is based on the theory that most numerical errors arise from large atomic condition numbers. While this theoretical foundation is well-supported and has proven effective in detecting significant errors in benchmarks such as ATOMU, HSED, and various linear system-solving programs, certain practical and complex programs may involve unpredictable conditions that could render our method ineffective.

%% file: relatedwork.tex
\section{Related Work}
\label{related-work}
In this section, we present research threads related to error detection in floating-point computations.

\subsection{Detecting floating-point exceptions}
Ariadne \cite{barr2013automatic} modifies a numerical program to include explicit checks for each condition that can trigger an exception, performing symbolic execution using real arithmetic to find real-valued inputs that could induce an exception.  
NUMFUZZ \cite{ma2022numfuzz} is a fuzzing tool that detects floating-point exceptions for numerical programs. It introduces a novel mutation strategy tailored for the floating-point format, designed to improve grey-box fuzzing by effectively generating valid floating-point test inputs. 

Our target numerical bugs are different from traditional exceptions. While exceptions like Overflow, Underflow, Inexact, Invalid, and Divide-By-Zero can typically be detected by compilers, certain bugs caused by large atomic condition numbers may only lead to inaccuracies, rather than triggering an exception. 
In cases where reliable oracles are unavailable, these inaccuracies can remain undetected.

\subsection{Conditioning}
To measure the inherent stability of a mathematical function, Higham et al.~\cite{higham2002accuracy} propose condition number, which could be derived using the Taylor series expansion. Since the condition number is hardly obtained for complex functions, study \cite{fu2015automated} presents an automated method based on Monte Carlo Markov Chain techniques to estimate condition numbers. Recently, Wang et al.~\cite{wang2022detecting} coarsely estimate condition numbers for real numerical programs and take them as fitnesses to guide search.

\subsection{Computing the oracles of floating-point programs}
Oracles are indispensable for many subjects, such as detecting errors in floating-point operations and searching for error-inducing inputs of programs. \textbf{1) With respect to detection:} FpDebug \cite{benz2012dynamic} is designed based on MPFR \cite{fousse2007mpfr} and Valgrind \cite{nethercote2007valgrind}. This tool analyzes numerical programs by conducting all operations side-by-side in high precision. \cite{chowdhary2021parallel} proposed a novel prototype for parallel shadow execution, named FPSanitizer, to find errors comprehensively on the multicore machines. EFTSanitizer \cite{chowdhary2022fast} further reduces the overheads by utilizing error-free transformation. \textbf{2) As for search:} Numerous tools \cite{chiang2014efficient, zou2015genetic, yi2019efficient, guo2020efficient, wang2022detecting, zhang2023eiffel, zhang2024hierarchical} have to obtain oracles to help them guide search. They tend to perform floating-point computations in higher precision by using libraries such as \textit{mpmath} \cite{mpmath}. 


To the best of our knowledge, ATOMU \cite{zou2019detecting} is the only tool that does not use high-precision program calculations to guide search. They consider numerical errors could be attributed to big condition numbers of atomic operations, such as catastrophic cancellation. They find inputs that result in huge condition number for each suspicious atomic operation by utilizing an evolutionary algorithm. However, some errors may be suppressed or masked during execution \cite{bao2013fly}, thus it is not necessary and time-consuming to detect big condition numbers for all operations. Our method injects perturbation for the result of each operation. Therefore, if ill-conditioned problems result in significant bugs in the output, our approach could report them.